\documentclass[journal= acsodf,manuscript=article]{achemso}
\usepackage{achemso}
\usepackage{graphicx}
\usepackage[utf8]{inputenc}
\usepackage{amsmath,amsfonts,amssymb}
\usepackage[usenames, dvipsnames]{color}
\usepackage{float}
\title{Electronic structure of the Sn$_{\text{1-x}}$Mn$_{\text{x}}$Te alloy: a resonant photoemission study}

\author{Elzbieta~Guziewicz}
\email{guzel@ifpan.edu.pl}
\affiliation{Institute of Physics, Polish Academy of Sciences, 02-668 Warsaw, Poland}

\author{Bronislaw~A. Orlowski}
\affiliation{Institute of Physics, Polish Academy of Sciences, 02-668 Warsaw, Poland}

\author{Bogdan~J. Kowalski}
\affiliation{Institute of Physics, Polish Academy of Sciences, 02-668 Warsaw, Poland}

\title{Electronic structure of Sn$_{\text{1-x}}$Mn$_{\text{x}}$Te semiconducting solid solution: a resonant photoemission study}

\begin{document}

\begin{abstract}

Manganese-doped tin telluride, Sn$_{\text{1-x}}$Mn$_{\text{x}}$Te, initially investigated as diluted magnetic semiconductor, has recently attracted considerable attention as a prospective thermoelectric material. The introduction of Mn was found to modify the valence band electronic structure, resulting in an improvement in the Seebeck coefficient and thus, the figure of merit (ZT). In the paper, we present a synchrotron radiation study of the electronic band structure of Sn$_{\text{0.9}}$Mn$_{\text{0.1}}$Te by resonant photoemission. The contribution of the $Mn3d$ electrons to the valence band (VB), calculated as the difference between the Energy Distribution Curves (EDCs) taken at the maximum and minimum of the Fano resonance for the $Mn3p$ - $Mn3d$ absorption threshold, shows a contribution at the VB edge, a dominant maximum at 4 eV, as well as a wide structure between 7 and 11 eV. 
Moreover, comparison with undoped SnTe reveals strong renormalization of the $Sn5p$ and $Te5p$ electronic states, which certainly influences the shape of the upper part of the valence band and electron effective mass.
\end{abstract}

\newpage
\section{Introduction}
Tin telluride is a IV-VI semiconductor with a direct and narrow energy gap (E$_{\text{g}}$ = 0.18 - 0.3 eV). It has been investigated since the 1960s due to the simple rock-salt crystallographic structure and narrow bandgap, which enables its use in optoelectronic devices operating in the infrared range. In 2012, SnTe was identified as one of the first topological crystalline insulators (TCI) when it was discovered that its surface supports topologically protected electronic states, i.e., resulting from mirror symmetry of its crystallographic structure.\cite{Hsieh2012, Tanaka2012} This discovery led to a fundamental shift in the understanding of material topology, opened a new branch of solid-state physics, and focused considerable attention on SnTe, which has become a model material for studying the influence of strain, pressure and doping on topological effects.\cite{Sadowski2018}
In parallel, starting from 1960s, tin telluride was also investigated as a lead-free thermoelectric material. However, its ZT figure of merit was found to be strongly limited mainly due to the high hole concentration (above 10$^{21}$/cm$^{3}$), too narrow E$_{\text{g}}$ and a large separation between light- and heavy-hole bands (0.3 eV versus 0.17 eV for PbTe) as well as a high lattice thermal conductivity (about 3.5 W/mK versus 1.5 W/mK for PbTe) that lead to the figure of merit, ZT, at the level of 0.3.\cite{Zhou2014}  The latter properties result in a low Seebeck coefficient compared to PbTe. For this reason, attempts have been made to model the electrical and thermal transport properties of SnTe by doping. In the first step, In dopant has been tested to modify the SnTe electronic structure by creating resonant levels in the valence band.\cite{Zhang2013} However, in 2015, $Tan$ $et$ $al.$ reported that Mn alloying in SnTe could also enhance its thermoelectric properties, in similar way to Mg, In, Hg and Cd alloying.\cite{Tan2015} This solution attracted a considerable scientific interest, as the ZT figure of merit was reported to be enhanced to 1.3 and several theoretical Density Functional Theory (DFT) calculations has been performed to establish the electronic band structure of the Sn$_{\text{1-x}}$Mn$_{\text{x}}$Te ternary alloy.\cite{Wu2015} Experimental photoemission studies of this system are limited to the simple determination of the $Mn3d$ contribution to the valence band \cite{Nadolny1998} and to the study of the deep core levels of the $Sn3d, Te3d$ and the $Mn2p$ orbitals \cite{Orlowski2002}.
In this paper, we present the synchrotron radiation photoemission study of Sn$_{\text{0.9}}$Mn$_{\text{0.1}}$Te and SnTe semiconductors.The photoemission spectra were taken in the  Constant Initial State (CIS) and Energy Distribution Curve (EDC) modes. Comparison with SnTe allows us to draw conclusions about the hybridization of the host crystal states with the manganese electrons.

\section{Experimental conditions}
Bulk SnTe and Sn$_{\text{0.9}}$Mn$_{\text{0.1}}$Te crystals were grown by a modified Bridgman method at the Institute of Physics, Polish Academy of Sciences. The photoemission experiment was performed on the FLIPPER II beamline in HASYLAB, Hamburg, Germany. Prior to the experiment, the sample was placed in a preparation chamber under ultra-high-vacuum conditions (UHV, p = $8\cdot 10^{-8}$ Pa) and clean surface was obtained by \emph {in situ} filing with a diamond file. A cylindrical mirror electron energy analyser was employed to measure the photoelectron spectra (PES) over a 47-60 eV photon energy range with an energy resolution of 0.2 eV. PES spectra were collected in two modes, Constant Initial State (CIS), in order to confirm the Fano-type shape of the resonance, and a typical photoemission spectra known as Energy Distribution Curve (EDC) that can be compared to density of electronic states. All the spectra presented in the paper are normalized to the photon flux and a background from the secondary electrons has been calculated by the Shirley method and subtracted. The binding energy (BE) scale is referred to the Fermi level measured for a metal foil in contact with the sample.

\section{Results and Discussion}
Resonant photoemission (RESPES) is a very convenient tool for studying the modification of the valence band by the $Mn3d$ electrons. This method requires a tunable light source provided by synchrotrons, because the energy of incident photons should be adjusted to the threshold of the $Mn3p-Mn3d$ intra-shell transition. In such a case, the classical photoexcitation process from the $Mn3d$ electron shell is accompanied by the resonant one, in which electron from the $Mn3p$ state is excited to the $Mn3d$ state. The latter excitation is followed by super-Coster-Kronig decay, when excited electron fills the $Mn3p$ hole, while its energy is transferred to another $Mn3d$ electron (Fig. 1a). As a result of quantum interference between two photoemission paths with the same initial and final states, we observe a strong and rapid variation of photoemission intensity from the $Mn3d$ electron shell. The phenomenon is based on the Fano effect \cite{Fano1961} and is sucessfully used for investigation of a contribution of the $3d, 4f$ and $5f$ electrons to the electronic band structure of a material \cite{Guziewicz2006}. The PES intensity has a charactristic Fano-type shape with a resonance maximum that appears just above the resonance energy and an antiresonance minimum that is observed at a slightly lower energy (Fig. 1b). It is assumed that at an antiresonance energy both, classical and resonant, transitions are opposite in phase and for aniresonance excitation we do not observe any contribution from the $Mn3d$ electrons. Therefore, the difference between resonant and antiresonant EDCs, called $\Delta$EDC, is assumed to reveal the contribution of the $Mn3d$ electrons to the valence band density of states.

\begin{figure}
   \centering
   \includegraphics[width =0.9\linewidth]{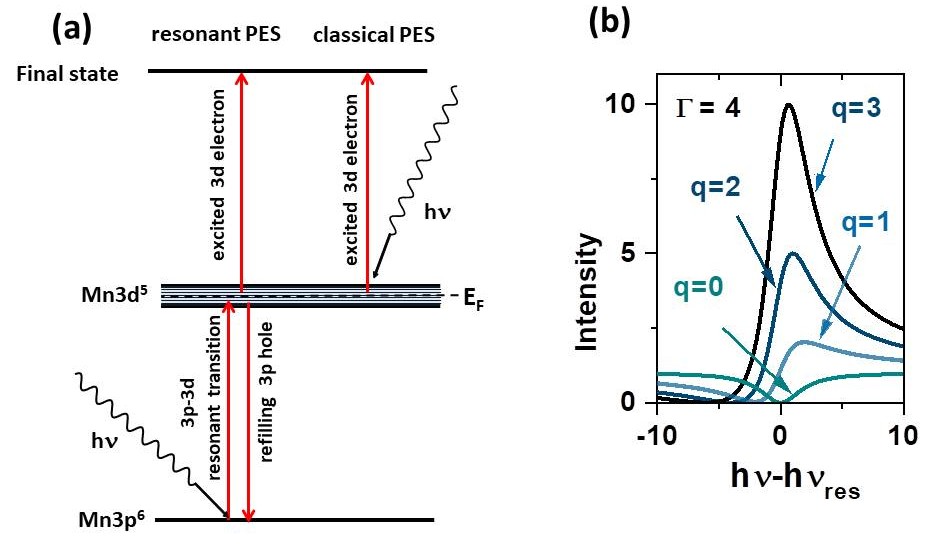}
    \caption{(a) The idea of the resonant photoemission; (b) The shape of the Fano resonance. 
  }
    \label{fig:Fig 1}
\end{figure}

Manganese is a transition metal element with five electrons on the $Mn3d$ electron shell, and the classical photoemission excitation process of electrons from the $Mn3d$ state can be expressed as follows:
\begin{equation}\label{clasicPE} 
Mn\ 3p^{6}3d^{5} + h\nu \rightarrow Mn\ 3p^{6}3d^{4} + e^-                       
\end{equation}
 When excitation energy approaches the $Mn3p-3d$ transition, which for Mn appears around 50 eV, the classical photoemission is accompanied by the resonant excitation  followed by super-Coster-Kronig decay:
 
\begin{equation}\label{resonancePE}
Mn\ 3p^{6}3d^{5} + h\nu \rightarrow Mn\ 3p^{5}3d^{6*} \rightarrow Mn\ 3p^{6}3d^{4} + e^-
\end{equation}

Fig. 2a shows a set of EDSs of Sn$_{\text{0.9}}$Mn$_{\text{0.1}}$Te crystal acquired for the photon energies between 47 and 52 eV, i.e. covering the energy range of the $Mn\ 3p \rightarrow 3d$ absorption threshold \cite{Kaurila1997}. Starting from h$\nu$ = 49 eV, we observe a significant increase in the photoemission intensity at BE of about 7 eV, and for h$\nu$ = 50 eV a new peak appears at BE of 4 eV, which becomes enhanced for h$\nu$ = 51 eV (see black arrows in Fig. \ref{fig:Fig 1}a). Comparison with the photoemission spectra of SnTe reveals that no such enhancement is observed for corresponding parts of EDCs for analogous photon energy (Fig. \ref{fig:Fig 1}b).

\begin{figure}
   \centering
   \includegraphics[width =0.8\linewidth]{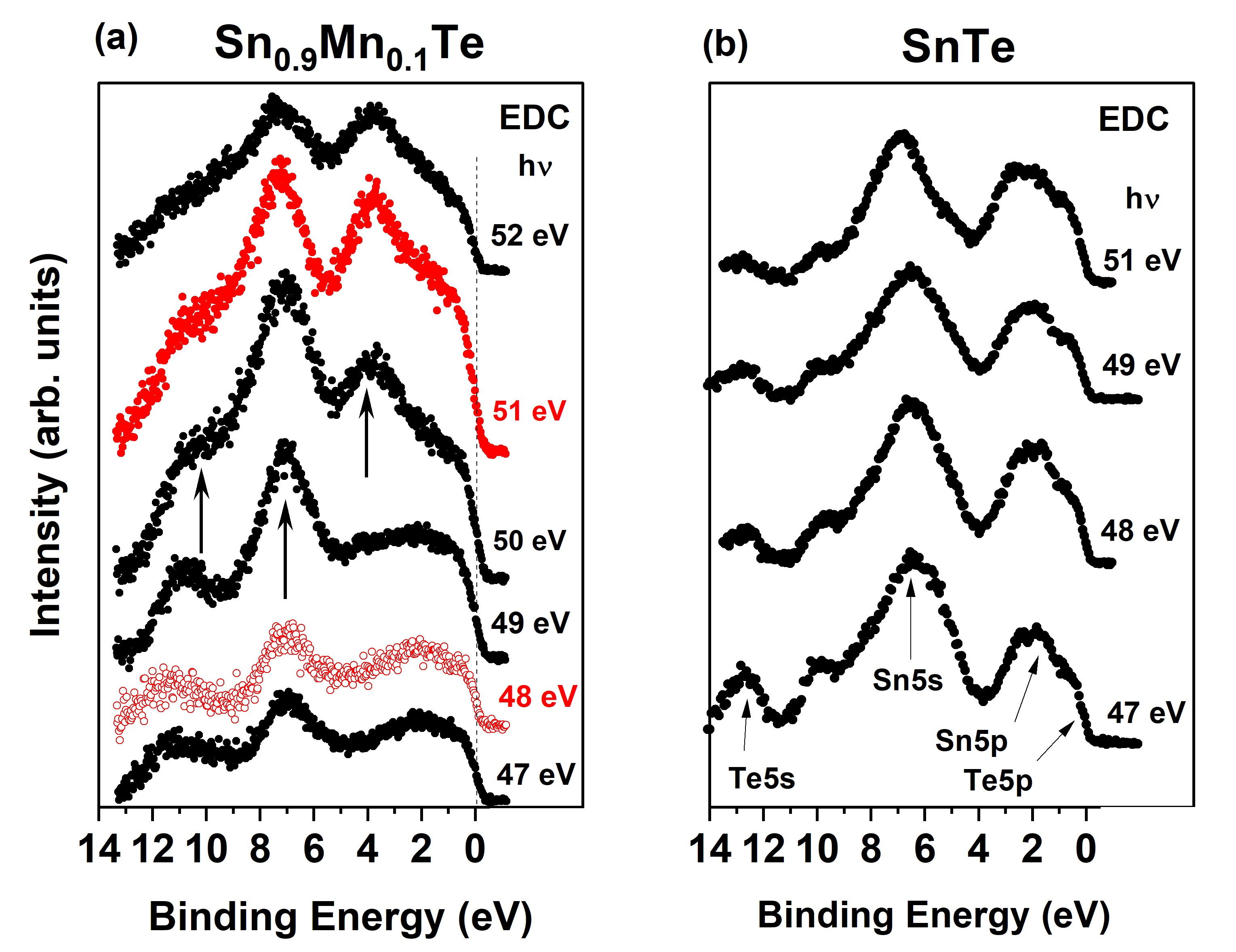}
    \caption{(a) A set of EDC spectra of Sn$_{\text{0.9}}$Mn$_{\text{0.1}}$Te measured for photon energy 47 - 52 eV; (b) A similar set of EDC spectra of the SnTe crystal. All the spectra are vertically shifted for clarity. 
  }
    \label{fig:Fig 1}
\end{figure}

The valence band EDCs of SnTe originate from the hybridized states of $Sn5p$ and $Te5p$ (BE between 0 and 4 eV), the $Sn5p$ states located between 6 and 8 eV, and the $Te5s$ electrons situated around 12-13 eV.\cite{Kemeny1976} According to the data tables,\cite{Yeh1985} the $Sn5p, Sn5s, Te5p$ and $Te5s$ electronic states show only a minor photon energy dependence of the photoionization cross-sections for the photon energy range of 47 - 51 eV, which explains the similarity of all spectra presented in Fig. 2b. This proves that the significant change in the photoemission spectra shown in Fig. 2a is related to the presence of manganese.

In order to precisely determine the resonance and anti-resonance energy, a photoemission measurements were performed in the CIS mode (Fig. 3a), in which the photon energies and the detection range of the photoelectron kinetic energy analyzer are changed by the same energy value, which allows recording the photoemission intensity as a function of energy for the selected initial state.
The CIS spectra in Fig. 3b show full Fano-like profiles as measured in the photon energy range of 47 – 58 eV, i.e. across the $Mn\ 3p \rightarrow Mn\ 3d$ photoionization threshold, for the initial state binding energies of 4, 7.5 and 9 eV, as indicated in the EDC spectra in Fig. 2a. The numerical analysis of the CIS spectra enabled us to find the photon energies corresponding to the maximum ($h\nu = 51\ eV$) and minimum ($h\nu = 48\ eV$) intensity of the Fano-like curve. The subtraction of the EDCs taken at those energies gives us the difference spectrum visualizing the $Mn\ 3d$ contribution to the photoemission spectra (Fig. 4a).

\begin{figure}
   \centering
   \includegraphics[width =0.8\linewidth]{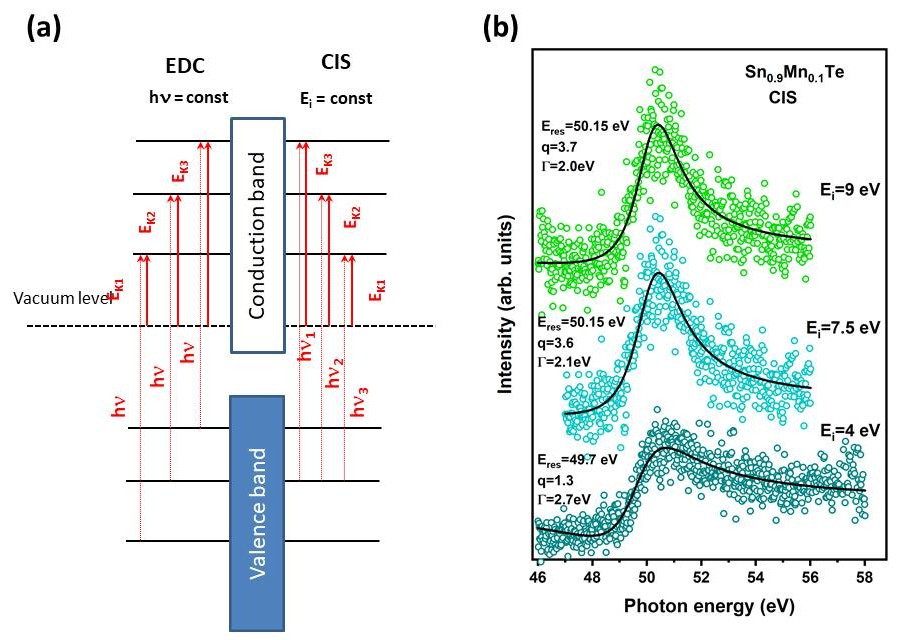}
    \caption{(a) The idea of two photoemission measurement modes: Energy Distribution Curve (EDC) and Constant Initial State (CIS); (b) CIS spectra of the Sn$_{\text{0.9}}$Mn$_{\text{0.1}}$Te crystal measured for photon energies near the $Mn3p - Mn3d$ photoexcitation threshold for initial energies 4, 7.5 and 9 eV below the Fermi level. The spectra are vertically shifted for clarity. 
  }
    \label{fig:Fig 2}
\end{figure}

Two main maxima, at the binding energies of 4 and 7.5 eV, and two shoulders, at about 1 and 9 eV, appear in the spectrum. The overall shape of the difference spectrum is consistent with a model spectrum obtained by a configuration interaction (CI) calculation on an octahedrally coordinated cluster Mn$^{\text{2+}}$(Te$^{\text{2-}})_{\text{6}}$.\cite{Ueda1994} The CI model calculation showed that the part of the spectrum in the top 6 eV was mainly due to transitions into the d$^{\text{5}}$\underline L final state configuration with a weak contribution of the d$^{\text{4}}$ configuration (\underline L represents a ligand hole). The part appearing for higher binding energies corresponds mainly to the transitions into the d$^{\text{4}}$ final states, without significant L$\rightarrow$d screening. Therefore, the part of the spectrum observed for higher binding energies can be interpreted as a satellite multiplet due to transitions to unscreened Mn 3d$^{\text{4}}$ final states (like for both MnTe \cite{Ueda1994} and Cd$_{\text{1-x}}$Mn$_{\text{x}}$Te with tetrahedral coordination of Mn ions \cite{Ley1986}). The low-binding-energy part of the spectrum, due to transitions to the final states screened by charge transfer from the valence band states (enabled by a considerable Mn 3d — Te 5p hybridization) can be interpreted as a measure of the density of the valence band states.

These conclusions can be supported by the analysis of the CIS spectra (Fig. \ref{fig:Fig 2}). The experimental curves are satisfactorily fitted with single Fano profiles \cite{Fano1961}

\begin{equation}\label{Fano}
I=I_0\frac{(\epsilon +q)^2}{(1+\epsilon ^2)}+I_{nres}
\end{equation}

\noindent
where $\epsilon = (h\nu -h\nu_{res})/(\Gamma$/2), $I_0$ describes the Mn 3d emission in the absence of the Fano autoionization, $I_{nres}$ is a non-interfering background contribution. The asymmetry parameter $q$ depends on the transition and interaction matrix elements. $I_0$ , $E_{res}$, $q$ and $\Gamma $ were used as the fit parameters. $I_{nres}$ was simulated by a linear function. The values for the fit parameters derived for the CIS spectra shown in Fig. \ref{fig:Fig 2} are listed in Table \ref{table:1}. 
\begin{table}
\begin{center}
\begin{tabular}{||c | c c c||} 
 \hline
 $E_i$ (eV) & $E_{res}$ (eV) & $q$ & $\Gamma $ (eV) \\ [0.5ex] 
 \hline\hline
 4.0 & 49.7 & 1.3 & 2.7 \\ 
 \hline
 7.5 & 50.15 & 3.6 & 2.1 \\
 \hline
 9.0 & 50.15 & 3.7 & 2.0 \\
 \hline
\end{tabular}
\end{center}
\caption{The values for the Fano profile parameters derived by fitting the CIS spectra of Sn$_{\text{0.9}}$Mn$_{\text{0.1}}$Te.}
\label{table:1}
\end{table}

The values for the Fano profile fitting parameters clearly fall into two groups - for the peak at $E_i$=4.0 eV and for two other features. The asymmetry parameter $q$ is much lower for the feature of $E_i$=4.0 eV. Its decrease can be ascribed to the presence of a larger
contribution from the transitions to the d$^{\text{n}}$\underline L final states \cite{Davis1982}. For this peak, the resonance energy becomes smaller by about 0.5 eV and the width of the Fano resonance (described by $\Gamma $) increases. 

  \begin{figure}
   \centering
   \includegraphics[width =0.9\linewidth]{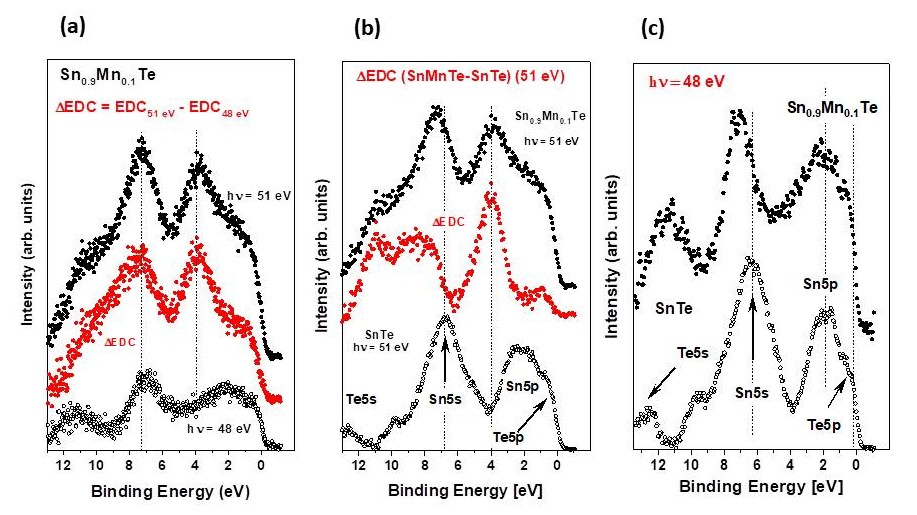}
    \caption{Differential spectra ($\Delta$EDC) showing the modification of the valence band by the $Mn3d$ electrons; (a) $\Delta$EDC as a difference between the resonant (h$\nu$ = 51 eV) and antiresonant (h$\nu$ = 48 eV) EDCs of the Sn$_{\text{0.9}}$Mn$_{\text{0.1}}$Te crystal; (b) $\Delta$EDC as a difference between EDCs of Sn$_{\text{0.9}}$Mn$_{\text{0.1}}$Te and SnTe crystals taken at the resonant energy (h$\nu$ = 51 eV); (c) $\Delta$EDC as a differnece between EDCs of Sn$_{\text{0.9}}$Mn$_{\text{0.1}}$Te and SnTe taken at the anti-resonant energy (h$\nu$ = 48 eV). The spectra are vertically shifted for clarity. 
  }
    \label{fig:Fig 2}
\end{figure}
The Fano curves corresponding to the features at 7.5 and 9.0 eV, contributing to a satellite appearing due to transitions to unscreened Mn 3d$^{\text{4}}$ final states, are described by sets of very similar fitting parameters. This confirms the conclusion from the CI model calculations about the significantly different nature of the upper and lower parts of the spectrum, related to $p-d$ hybridization.

A deeper insight into the problem of modification of the valence band of the SnTe crystal by Mn orbitals can be obtained by comparing the EDC spectra of Sn$_{\text{0.9}}$Mn$_{\text{0.1}}$Te and SnTe semiconductors taken at h$\nu$ = 51 eV (Fig. 4b). The difference spectra ($\Delta$EDC) obtained from their subtraction shows a strong signal enhancement at the VB edge as well as at BE of 4 eV, so both of them can be assigned to the $Mn3d$ states. We also see a broad structure situated between 7 and 11 eV, which, according to CI calculations, can be interpreted as a satellite structure of multi-electron interactions origin. Additional information can be obtained by comparing the EDC spectra of SnTe and Sn$_{\text{0.9}}$Mn$_{\text{0.1}}$Te measured at anti-resonant energy (48 eV), where the contribution of the $Mn3d$ electrons is negligible (Fig. 4c). In principle, both EDCs are related solely to the density of states of SnTe in the valence band region, however, the latter shows hybridization of SnTe orbitals resulting from interaction with Mn orbitals. As can be seen, both spectra are clearly different. The peak originating from the $Sn5s$ states is shifted by 0.6 eV towards higher binding energy, i.e. from 6.4 to 7.0 eV, while the one associated with the $Sn5p$ states is wider and more diffuse. Moreover, the intensity of photoemission at the VB edge originating from the $Te5p$ states is clearly enhanced. A significant change is also observed for states located deeper in the spectrum originating from the $Te5s$ orbitals that were initially located at BE of 12.5 eV. In the case of Sn$_{\text{0.9}}$Mn$_{\text{0.1}}$Te semiconductor the photoemission structure related to these states is much broader and more intensive forming a wide peak located between 10 and 12 eV. Thus, the EDC spectra collected at antiresonant energy (48 eV) indicate that the introduction of manganese into the SnTe semiconductor induces a strong modification of electronic states in the entire valence band region. 
The observed significant renormalization of the $Sn5p$ and $Te5p$ electronic states can be expected to influence the shape of the upper part of the valence band and thus the effective electron mass.
This means that the thermodynamic response of Sn$_{\text{1-x}}$Mn$_{\text{x}}$Te cannot be considered solely in terms of electronic states close to the VB maximum, because the deeper $Mn3d$ states, through strong $p-d$ hybridization, lead to the renormalization of the SnTe valence band dispersion and to the modification of the effective electron mass and possible degeneracy of the hole bands.

\section{Summary}
To summarize, the electronic band structure of Sn$_{\text{0.9}}$Mn$_{\text{0.1}}$Te and SnTe semiconductors has been studied using synchrotron radiation photoemission experiment. The CIS spectra measured for photon energies close to the $Mn3p - Mn3d$ absorption threshold show a Fano-like shape with resonance maxiumum at 50-51 eV. A comparison of $\Delta$EDCs obtained as a subtraction of resonant and antiresonant EDC with that resulting from subtraction of EDC$_{SnTe}$ from EDC$_{Sn_{\text{0.9}}Mn_{\text{0.1}}Te}$ shows similar features located at the VB edge, at 4 and 7 eV below. 
A comparison of EDC spectra of SnTe and Sn$_{\text{0.9}}$Mn$_{\text{0.1}}$Te taken at the antiresonance photon energy gives a fingerprint of a strong modification of the host SnTe orbitals by manganese ions.
In the light of photoemission studies carried out across the entire valence band, the thermodynamic response of Sn$_{\text{1-x}}$Mn$_{\text{x}}$Te cannot be considered solely in terms of electronic states close to the VB maximum, because the deeper $Mn3d$ states, through strong $p-d$ hybridization, lead to the renormalization of the SnTe valence band dispersion and to the modification of the effective electron mass and possible degeneracy of the hole bands. 

The results presented above might be important for understanding the physical properties of  Sn$_{\text{1-x}}$Mn$_{\text{x}}$Te solid solutions. The electronic band structure of this system remains an important crystal feature since its topology determines the structure of non-trivial surface states.

\newpage

\bibliography{SnMnTe}

\end{document}